\documentclass[conference]{IEEEtran}
\IEEEoverridecommandlockouts
\usepackage{cite}
\usepackage{amsmath,amssymb,amsfonts}
\usepackage{algorithmic}
\usepackage{graphicx}
\usepackage{textcomp}
\usepackage{xcolor}
\usepackage{multirow}
\usepackage{verbatim}
\usepackage[super]{nth}
\def\BibTeX{{\rm B\kern-.05em{\sc i\kern-.025em b}\kern-.08em
    T\kern-.1667em\lower.7ex\hbox{E}\kern-.125emX}}
\begin{document}

\title{Changes in Power and Information Flow in Resting-state EEG by Working Memory Process 

\thanks{20xx IEEE. Personal use of this material is permitted. Permission
from IEEE must be obtained for all other uses, in any current or future media, including reprinting/republishing this material for advertising or promotional purposes, creating new collective works, for resale or redistribution to servers or lists, or reuse of any copyrighted component of this work in other works.}

\footnote{{\thanks{This work was partly supported by Institute for Information \& Communications Technology Promotion (IITP) grants funded by the Korea government(MSIT) (No. 2015-0-00185: Development of Intelligent Pattern Recognition Softwares for Ambulatory Brain-Computer Interface, No. 2017-0-00451: Development of BCI based Brain and Cognitive Computing Technology for Recognizing User’s Intentions using Deep Learning, No. 2019-0-00079: Artiﬁcial Intelligence Graduate School Program, Korea University, and No. 2021-0-02068: Artificial Intelligence Innovation Hub).}
}}
}

\author{\IEEEauthorblockN{Gi-Hwan Shin}
\IEEEauthorblockA{\textit{Dept. Brain and Cognitive Engineering} \\
\textit{Korea University} \\
Seoul, Republic of Korea \\
gh\_shin@korea.ac.kr}  \\
\and

\IEEEauthorblockN{Young-Seok Kweon}
\IEEEauthorblockA{\textit{Dept. Brain and Cognitive Engineering} \\
\textit{Korea University}\\
Seoul, Republic of Korea \\
youngseokkweon@korea.ac.kr}
\and

\IEEEauthorblockN{Heon-Gyu Kwak}
\IEEEauthorblockA{\textit{Dept. Artificial Intelligence} \\
\textit{Korea University} \\
Seoul, Republic of Korea \\
hg\_kwak@korea.ac.kr} 
}

\maketitle

\begin{abstract}
Many studies have analyzed working memory (WM) from electroencephalogram (EEG). However, little is known about changes in the brain neurodynamics among resting-state (RS) according to the WM process. Here, we identified frequency-specific power and information flow patterns among three RS EEG before and after WM encoding and WM retrieval. Our results demonstrated the difference in power and information flow among RS EEG in delta (1-3.5 Hz), alpha (8-13.5 Hz), and beta (14-29.5 Hz) bands. In particular, there was a marked increase in the alpha band after WM retrieval. In addition, we calculated the association between significant characteristics of RS EEG and WM performance, and interestingly, correlations were found only in the alpha band. These results suggest that RS EEG according to the WM process has a significant impact on the variability and WM performance of brain mechanisms in relation to cognitive function.
\end{abstract}

\begin{small}
\end{small}

\section{INTRODUCTION}
Working memory (WM) is the ability to maintain and manipulate information over a short period of time, consisting of encoding, retention, and retrieval processes \cite{small2001circuit}. In the field of neuroscience, there is great interest in the neural oscillation dynamics of WM process using electroencephalogram (EEG) \cite{shin2021predicting}. EEG can extract rapidly fluctuating brain activation characteristics based on high temporal resolution \cite{won2017motion, lee2018high, kwon2019subject}. In general, many studies identify brain mechanisms during WM encoding and WM retrieval \cite{kragel2017similar}, but resting-state (RS) without stimulated neural activity is also receiving attention \cite{heister2013resting, pyka2009impact, zhang2017hybrid}. However, there are only a few studies on the effects of RS EEG in relation to the WM process.

The WM process can be analyzed using the amplitude and phase of the brain activity measured from the EEG \cite{shin2022differential}. Power spectral density (PSD) is a method of quantifying amplitude by calculating various frequency bands (delta, theta, alpha, beta, and gamma) using a fast Fourier transform (FFT) \cite{lee2019connectivity,lee2020frontal}. Among them, alpha bands are highly correlated with various cognitive functions, such as performance and information processing speed \cite{brokaw2016resting}. Phase transfer entropy (PTE) is a phase-based information flow estimation method for computing interactions between complex cortical \cite{lobier2014phase,ahmadi2020decoding}. Using this, several studies have reported the direction of information flow in brain networks in various frequency bands during WM \cite{hillebrand2016direction, wang2019consistency}. However, no studies have investigated the difference in power and information flow among RS EEG in relation to the WM process.

In this study, EEG data were obtained from twenty-nine participants to investigate changes in the neurodynamics in RS EEG according to the WM process. Participants performed a WM task and three RS EEG. We calculated the power and information flow among RS EEG before and after WM encoding and WM retrieval. We hypothesized that the power and information flow of frequencies in RS EEG will be greater as the WM task progressed. Also, we thought that there was a correlation between the significant EEG characteristics in RS and WM performance.

\begin{figure*}[t!]
\centering
\scriptsize
\includegraphics[width=\textwidth]{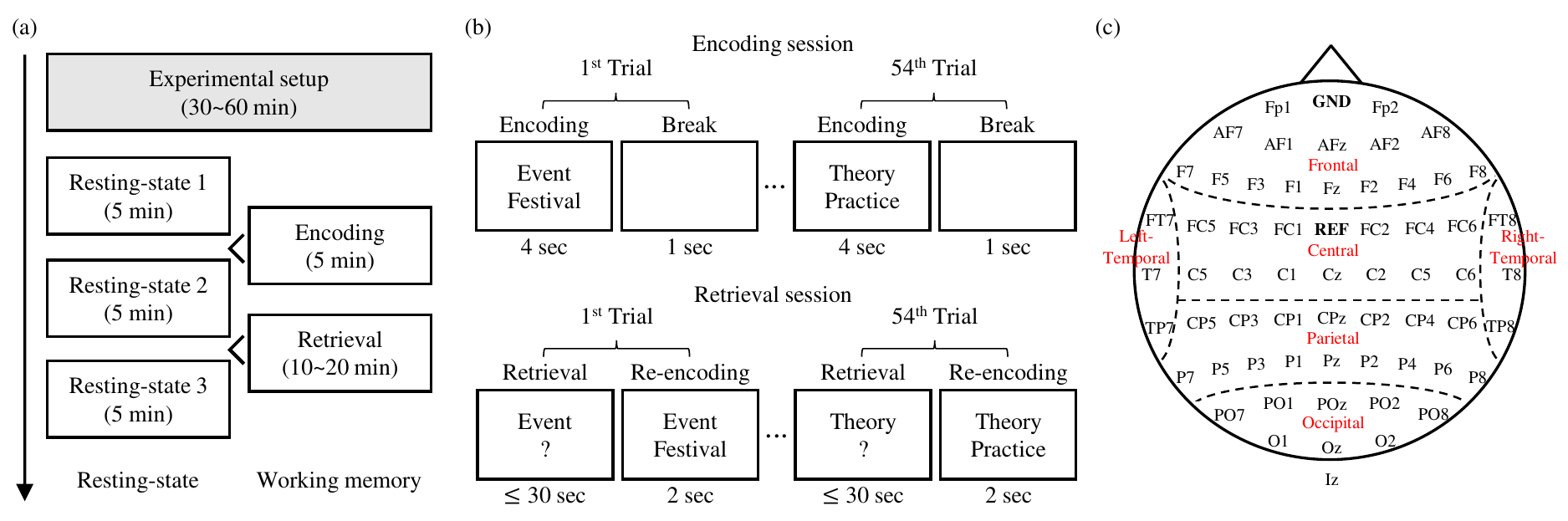}
\caption{Experimental design. (a) Experimental procedures involving resting-state and working memory. (b) Working memory task consisting of encoding and retrieval sessions. (c) Channel placement of 60 EEG electrodes and six regions of interest (frontal, central, left temporal, right temporal, parietal, and occipital regions.)}
\end{figure*}

\section{METHODS}
\subsection{Participants and Experimental Produce}
Twenty-nine participants (17 females; mean $\pm$ SD age: 25.1 $\pm$ 2.6 years) were recruited and participated in the study after obtaining written informed consent. They were free of any neurologic or psychiatric disorders. This study was approved by the Institutional Review Board at Korea University (KUIRB-2021-0155-03). 

This experiment consists of a WM task and RS EEG (Fig. 1a). First, each participant visited the laboratory and prepared for the experiment for about an hour. Then, a total of three 5 minutes of eye-closed RS were performed before and after WM process. WM performed 54 word-pair tasks (Fig. 1b) \cite{marshall2006boosting, shin2020assessment}. The encoding session displays word pairs for 4 seconds and breaks for 1 second. The retrieval session uses the keyboard to enter a pair of words displayed on the screen within 30 seconds. After that, the word pair is re-encoded by displaying the correct answer for 2 seconds. Memory performance evaluation was considered correct for typos and inflectional errors. The task was implemented with Psychtoolbox (http://psychtoolbox.org).

\subsection{Data Recording and Preprocessing}
The data recorded 60 EEG and 4 EOG signals from 64 Ag/AgCl electrodes using BrainAmp (Brain products GmBH, Germany) at a sampling rate of 1,000 Hz. The EEG electrodes were arranged based on the 10-20 international system configuration, and the EOG electrodes were arranged with two at both ends of the eye (horizontal) and two in the right eye (vertical). FCz and Fpz were used as reference and ground. The impedance of all electrodes was kept below 20 k$\Omega$.

The recorded EEG signals were preprocessed using the EEGLAB \cite{delorme2004eeglab} and BCILAB toolboxes \cite{kothe2013bcilab} for MATLAB 2018b. First, down-sampling was performed at 250 Hz and band-pass filtering was performed at 0.5 to 100 Hz. After that, three RS EEG were segmented and independent component analysis was applied to remove eye movements based on EOG signals \cite{jeong2020brain}. Finally, a Laplacian spatial filter was used to improve the signal-to-noise ratio \cite{jeong2020decoding}.

\subsection{EEG Data Analysis}
To identify the power and information flow among three RS EEG according to the WM process, we calculated PSD and PTE in five frequency bands: delta (1.3-5 Hz), theta (4-7.5 Hz), alpha (8-13.5 Hz), beta (14-29.5 Hz), and gamma (30-50 Hz). We also compared EEG characteristics between brain regions by grouping 60 channels into six brain regions of interest (ROIs): frontal, central, left temporal, right temporal, parietal, and occipital regions (Fig. 1c).

\subsubsection{Power Spectral Density}
To identify differences in the power distribution of RS EEG, the cleaned EEG signal was transformed into the frequency domain using FFT \cite{suk2014predicting}. PSD was calculated for each frequency band for all EEG channels.

\subsubsection{Phase Transfer Entropy}
PTE is a functional connectivity estimation method that can measure large-scale phase-specific directed connectivity between EEG channels using phase time-series data extracted from EEG \cite{hillebrand2016direction}. The dPTE calculation is detailed in Wang \textit{et al.} \cite{wang2019consistency}. For each participant's all EEG channels, the dPTE was calculated and grouped into six ROIs.

\subsection{Statistical Analysis}
To confirm the difference among RS EEG according to the WM process, we performed statistical verification for three conditions (RS 1 vs. RS 2, RS 1 vs. RS 3, and RS 2 vs. RS 3). First, a comparative analysis of PSD was performed using paired \textit{t}-test. Second, PTE was compared using a non-parametric permutation test (\textit{r} = 5,000), considering that not all PTE values were normally distributed from the Lilliefors test. Finally, the Pearson correlation coefficient was calculated to determine the relationship between the significant EEG characteristics and WM performance. The \textit{p}-values for all analyses are 0.01.

\begin{figure*}[t!]
\centering
\scriptsize
\includegraphics[width=\textwidth]{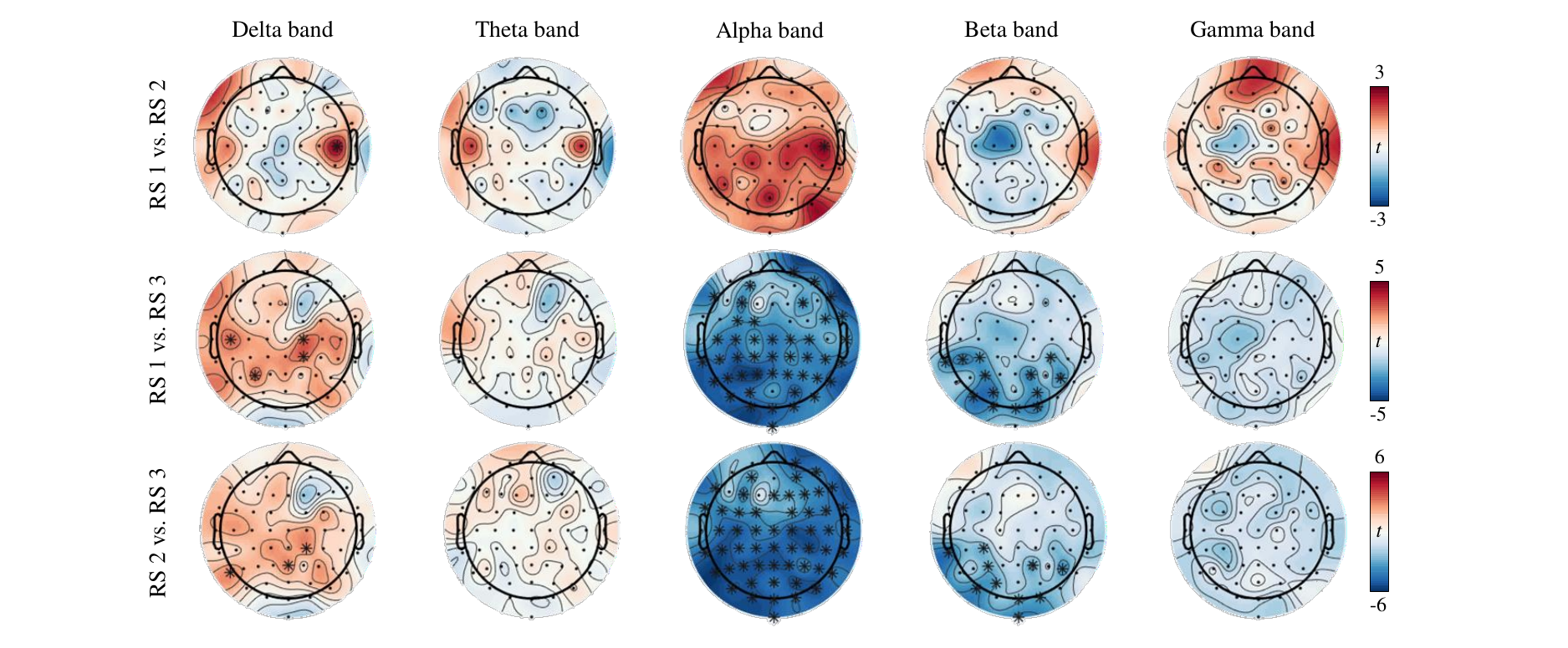}
\caption{Statistical differences in spectral power among RS EEG in each frequency band. Each color bar represents the \textit{t}-values. The black asterisk indicates a significant channel (\textit{p} $<$ 0.01).}
\end{figure*}

\begin{figure*}[t!]
\centering
\scriptsize
\includegraphics[width=\textwidth]{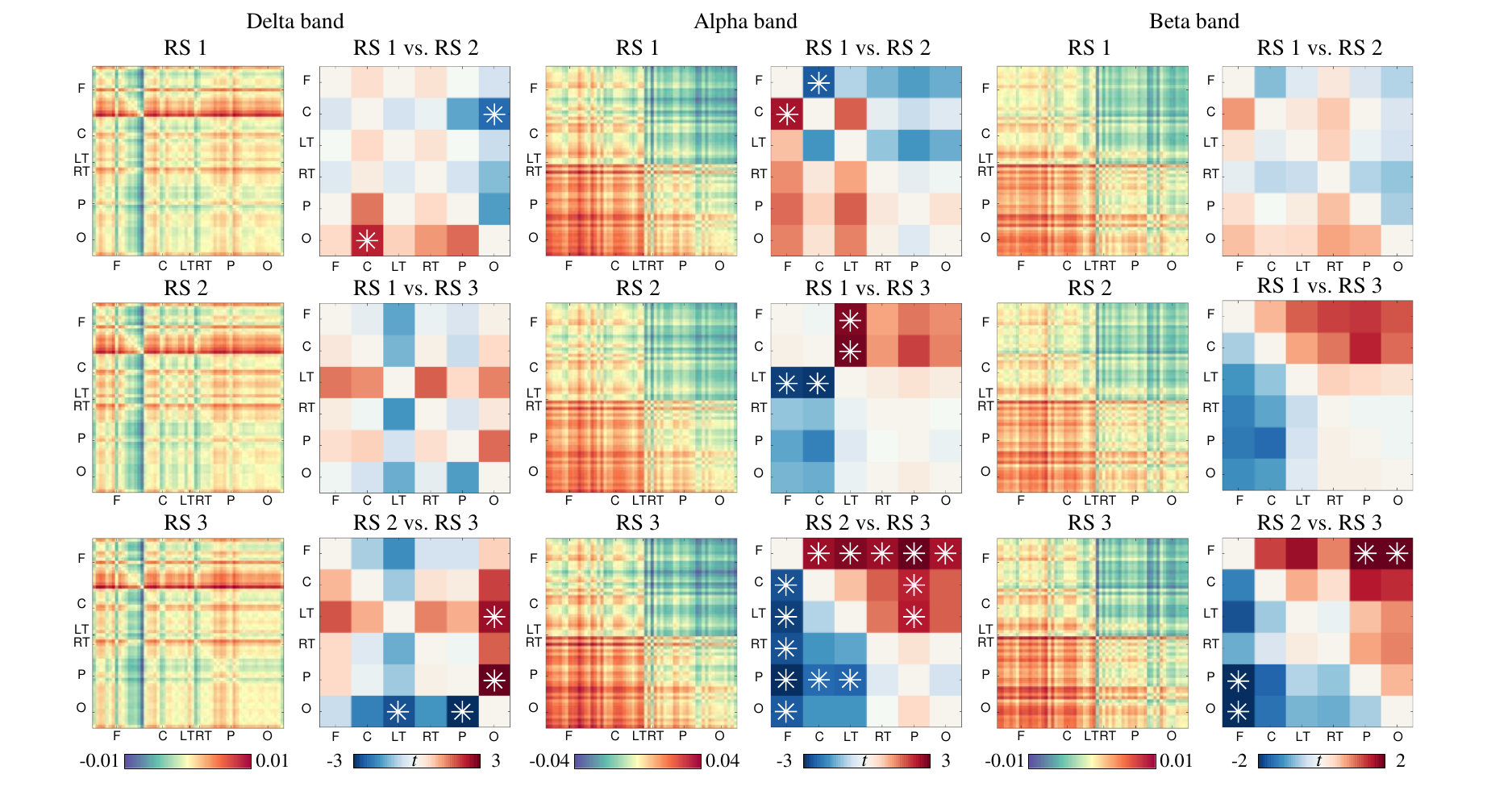}
\caption{Mean dPTE and statistical difference for RS EEG in delta, alpha, and beta bands. Each color bar represents the dPTE and \textit{t}-values. The white asterisk indicates significant connectivity (\textit{p} $<$ 0.01).}
\end{figure*}

\begin{figure*}[t!]
\centering
\scriptsize
\includegraphics[width=\textwidth]{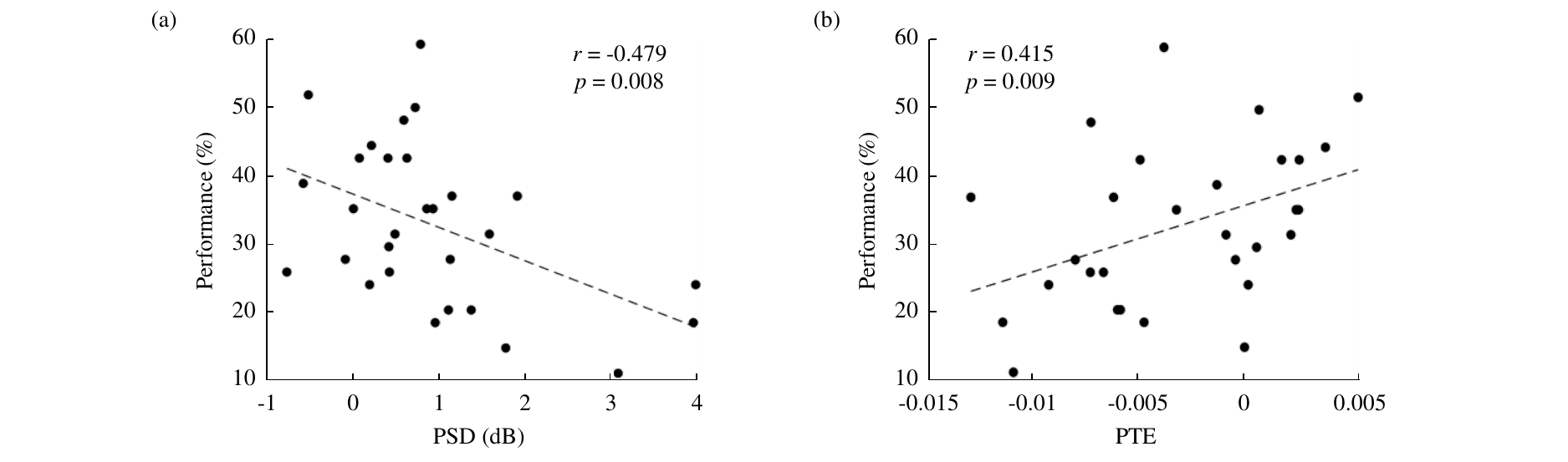}
\caption{Correlation between significant EEG characteristics of the alpha band and WM performance. (a) and (b) represent PSD and PTE, respectively.}
\end{figure*}

\section{RESULTS}
\subsection{Changes in Spectral Power of RS EEG}
We calculated spectral power to identify changes in RS EEG according to the WM process. Fig. 2 shows the statistical differences among RS by each frequency band. In the delta band, there was a significant difference in the channel near the central and parietal regions. In the alpha band, it was confirmed that brain activation was statistically higher as WM process progressed. In the beta band, RS 1 vs. RS 2 and RS 2 vs. RS 3 showed differences in channels near the parietal and occipital regions. Conversely, there was no statistical difference in theta and gamma bands. 

\subsection{Difference in Directional Information Flow among RS EEG}
To investigate the pattern changes of directional information flow according to the WM load, we analyzed RS EEG using dPTE. Fig. 3 shows the results of information flow in the delta, alpha, and beta bands. The dPTE values between channels for each RS showed an anterior-to-posterior information flow in the delta band, and conversely, a posterior-to-anterior information flow in the alpha and beta bands. The \textit{t}-value of dPTE between ROIs of RS EEG was most prominent in RS 2 vs. RS 3, and in particular, the posterior-to-anterior information flow of the alpha band was statistically higher in RS 3 than in RS 2. On the other hand, information flow was dispersed or weak in theta and gamma bands.

\subsection{Correlation between Alpha Band and WM Performance}
We investigated the effect of RS EEG before and after WM encoding and WM retrieval on WM performance. The results showed a relationship with WM performance only in the difference between RS 3 and RS 1 in the alpha band (Fig. 4). Specifically, Fig. 4a shows a negative correlation between spectral power in the left temporal region and performance and Fig. 4b shows a positive correlation between information flow from left temporal to frontal regions and performance. On the other hand, other EEG characteristics did not show a significant relationship with performance.

\section{DISCUSSION}
In the current study, we identified the differences in spectral power in delta, alpha, and beta bands among RS EEG before and after WM encoding and WM retrieval. The previous study has reported that changes in the frequencies of RS EEG reflect state transitions of brain activity during WM \cite{lopez2013alterations}. This suggests that RS EEG may be modulated by the demands of cognitive tasks. Our results provide evidence that brain activation changes in several frequencies in RS EEG are influenced by previous WM tasks.

Interestingly, we found that the information flow of channels and ROIs among RS EEG became much stronger at certain frequencies after WM retrieval. The anterior-to-posterior information flow in the delta band and the posterior-to-anterior information flow in the alpha and beta bands were consistent with the results of previous studies \cite{massimini2004sleep, hillebrand2016direction, wang2019consistency}. Information flow has been reported to be much stronger in the alpha band associated with internal mental synchronization that inhibits the processing of previously incoming visual stimuli \cite{scheeringa2012eeg}. In addition, previous functional magnetic resonance imaging studies have reported that a challenging WM task has a significant effect on the activation of the default mode network during subsequent RS \cite{forbes2015spontaneous, pyka2009impact}. Taken together, dynamic changes in the information flow in delta, alpha, and beta bands during RS EEG can help to explain their functional role in cognitive neural processing.

We found correlations between the alpha band of RS EEG and WM performance. The spectral power is negatively correlated with WM performance as an active integration mechanism similar to that occurring during sleep \cite{brokaw2016resting}. In addition, it may explain the notion that increased long-range coherence of information flows reflects the central executive function of the WM \cite{sauseng2005fronto}. These findings suggest that power and information flow in the alpha band during RS EEG are important markers related to WM performance.

As a limitation of this study, the conclusion was drawn only for changes in RS EEG according to verbal WM. Therefore, we will further investigate the variability among RS EEG by performing various WM tasks such as N-back \cite{heister2013resting} or visuospatial tasks \cite{shin2020assessment}. In addition, we will further investigate the relationship between RS EEG and WM process by analyzing the interactions among significant frequencies \cite{thung2018conversion, kim2019subject}.

In conclusion, our results showed that RS EEG according to the WM process had a significant impact on the variability of brain mechanisms and performance in relation to cognitive function. In particular, the changes in power and information flow in the alpha band were most prominent in RS EEG after WM retrieval. Therefore, these findings suggest that RS EEG may be useful for understanding cognitive neuroscience.


\bibliographystyle{IEEEtran}
\bibliography{REFERENCE}

\begin{thebibliography}{10}
\providecommand{\url}[1]{#1}
\csname url@samestyle\endcsname
\providecommand{\newblock}{\relax}
\providecommand{\bibinfo}[2]{#2}
\providecommand{\BIBentrySTDinterwordspacing}{\spaceskip=0pt\relax}
\providecommand{\BIBentryALTinterwordstretchfactor}{4}
\providecommand{\BIBentryALTinterwordspacing}{\spaceskip=\fontdimen2\font plus
\BIBentryALTinterwordstretchfactor\fontdimen3\font minus
  \fontdimen4\font\relax}
\providecommand{\BIBforeignlanguage}[2]{{%
\expandafter\ifx\csname l@#1\endcsname\relax
\typeout{** WARNING: IEEEtran.bst: No hyphenation pattern has been}%
\typeout{** loaded for the language `#1'. Using the pattern for}%
\typeout{** the default language instead.}%
\else
\language=\csname l@#1\endcsname
\fi
#2}}
\providecommand{\BIBdecl}{\relax}
\BIBdecl

\bibitem{small2001circuit}
S.~A. Small \emph{et~al.}, ``Circuit mechanisms underlying memory encoding and
  retrieval in the long axis of the hippocampal formation,'' \emph{Nat.
  Neurosci.}, vol.~4, no.~4, pp. 442--449, 2001.

\bibitem{shin2021predicting}
G.-H. Shin, Y.-S. Kweon, and M.~Lee, ``Predicting the transition from
  short-term to long-term memory based on deep neural network,'' in \emph{Int.
  Winter Conf. Brain-Computer Interface (BCI)}.\hskip 1em plus 0.5em minus
  0.4em\relax Jeongseon, Republic of Korea, Feb., 2021, pp. 1--5.

\bibitem{won2017motion}
D.-O. Won, H.-J. Hwang, D.-M. Kim, K.-R. M{\"u}ller, and S.-W. Lee,
  ``Motion-based rapid serial visual presentation for gaze-independent
  brain-computer interfaces,'' \emph{IEEE Trans. Neural Syst. Rehabil. Eng.},
  vol.~26, no.~2, pp. 334--343, 2017.

\bibitem{lee2018high}
M.-H. Lee, J.~Williamson, D.-O. Won, S.~Fazli, and S.-W. Lee, ``A high
  performance spelling system based on {EEG}-{EOG} signals with visual
  feedback,'' \emph{IEEE Trans. Neural Syst. Rehabil. Eng.}, vol.~26, no.~7,
  pp. 1443--1459, 2018.

\bibitem{kwon2019subject}
O.-Y. Kwon, M.-H. Lee, C.~Guan, and S.-W. Lee, ``Subject-independent
  brain--computer interfaces based on deep convolutional neural networks,''
  \emph{IEEE Trans. Neural Netw. Learn. Syst.}, vol.~31, no.~10, pp.
  3839--3852, 2019.

\bibitem{kragel2017similar}
J.~E. Kragel \emph{et~al.}, ``Similar patterns of neural activity predict
  memory function during encoding and retrieval,'' \emph{Neuroimage}, vol. 155,
  pp. 60--71, 2017.

\bibitem{heister2013resting}
D.~Heister \emph{et~al.}, ``Resting-state neuronal oscillatory correlates of
  working memory performance,'' \emph{PLoS One}, vol.~8, no.~6, p. e66820,
  2013.

\bibitem{pyka2009impact}
M.~Pyka \emph{et~al.}, ``Impact of working memory load on f{MRI} resting state
  pattern in subsequent resting phases,'' \emph{PLoS One}, vol.~4, no.~9, p.
  e7198, 2009.

\bibitem{zhang2017hybrid}
Y.~Zhang, H.~Zhang, X.~Chen, S.-W. Lee, and D.~Shen, ``Hybrid high-order
  functional connectivity networks using resting-state functional {MRI} for
  mild cognitive impairment diagnosis,'' \emph{Sci. Rep.}, vol.~7, no.~1, pp.
  1--15, 2017.

\bibitem{shin2022differential}
G.-H. Shin and Y.-S. Kweon, ``Differential {EEG} characteristics during working
  memory encoding and re-encoding,'' in \emph{Int. Winter Conf. Brain-Computer
  Interface (BCI)}.\hskip 1em plus 0.5em minus 0.4em\relax Jeongseon, Republic
  of Korea, Feb., 2022, pp. 1--4.

\bibitem{lee2019connectivity}
M.~Lee \emph{et~al.}, ``Connectivity differences between consciousness and
  unconsciousness in non-rapid eye movement sleep: A {TMS}--{EEG} study,''
  \emph{Sci. Rep.}, vol.~9, no.~1, pp. 1--9, 2019.

\bibitem{lee2020frontal}
M.~Lee, G.-H. Shin, and S.-W. Lee, ``Frontal {EEG} asymmetry of emotion for the
  same auditory stimulus,'' \emph{IEEE Access}, vol.~8, pp. 107\,200--107\,213,
  2020.

\bibitem{brokaw2016resting}
K.~Brokaw \emph{et~al.}, ``Resting state {EEG} correlates of memory
  consolidation,'' \emph{Neurobiol. Learn. Mem.}, vol. 130, pp. 17--25, 2016.

\bibitem{lobier2014phase}
M.~Lobier, F.~Siebenh{\"u}hner, S.~Palva, and J.~M. Palva, ``Phase transfer
  entropy: a novel phase-based measure for directed connectivity in networks
  coupled by oscillatory interactions,'' \emph{Neuroimage}, vol.~85, pp.
  853--872, 2014.

\bibitem{ahmadi2020decoding}
A.~Ahmadi, S.~Davoudi, M.~Behroozi, and M.~R. Daliri, ``Decoding covert visual
  attention based on phase transfer entropy,'' \emph{Physiol. Behav.}, vol.
  222, p. 112932, 2020.

\bibitem{hillebrand2016direction}
A.~Hillebrand \emph{et~al.}, ``Direction of information flow in large-scale
  resting-state networks is frequency-dependent,'' \emph{Proc. Natl. Acad. Sci.
  U. S. A.}, vol. 113, no.~14, pp. 3867--3872, 2016.

\bibitem{wang2019consistency}
R.~Wang \emph{et~al.}, ``Consistency and dynamical changes of directional
  information flow in different brain states: A comparison of working memory
  and resting-state using {EEG},'' \emph{Neuroimage}, vol. 203, p. 116188,
  2019.

\bibitem{marshall2006boosting}
L.~Marshall, H.~Helgad{\'o}ttir, N.~M{\"o}lle, and J.~Born, ``Boosting slow
  oscillations during sleep potentiates memory,'' \emph{Nature}, vol. 444, no.
  7119, pp. 610--613, 2006.

\bibitem{shin2020assessment}
G.-H. Shin, M.~Lee, and S.-W. Lee, ``Assessment of unconsciousness for memory
  consolidation using {EEG} signals,'' in \emph{Conf. Proc. IEEE. Int. Conf.
  Syst. Man Cybern. (SMC)}.\hskip 1em plus 0.5em minus 0.4em\relax Toronto,
  Canada, Oct., 2020, pp. 513--519.

\bibitem{delorme2004eeglab}
A.~Delorme and S.~Makeig, ``{EEGLAB}: An open source toolbox for analysis of
  single-trial {EEG} dynamics including independent component analysis,''
  \emph{J. Neurosci. Methods}, vol. 134, no.~1, pp. 9--21, 2004.

\bibitem{kothe2013bcilab}
C.~A. Kothe and S.~Makeig, ``{BCILAB}: A platform for brain--computer interface
  development,'' \emph{J. Neural Eng.}, vol.~10, no.~5, p. 056014, 2013.

\bibitem{jeong2020brain}
J.-H. Jeong, K.-H. Shim, D.-J. Kim, and S.-W. Lee, ``Brain-controlled robotic
  arm system based on multi-directional {CNN}-{BiLSTM} network using {EEG}
  signals,'' \emph{IEEE Trans. Neural Syst. Rehabil. Eng.}, vol.~28, no.~5, pp.
  1226--1238, 2020.

\bibitem{jeong2020decoding}
J.-H. Jeong, N.-S. Kwak, C.~Guan, and S.-W. Lee, ``Decoding movement-related
  cortical potentials based on subject-dependent and section-wise spectral
  filtering,'' \emph{IEEE Trans. Neural Syst. Rehabil. Eng.}, vol.~28, no.~3,
  pp. 687--698, 2020.

\bibitem{suk2014predicting}
H.-I. Suk, S.~Fazli, J.~Mehnert, K.-R. M{\"u}ller, and S.-W. Lee, ``Predicting
  {BCI} subject performance using probabilistic spatio-temporal filters,''
  \emph{PLoS One}, vol.~9, no.~2, 2014.

\bibitem{lopez2013alterations}
R.~A. L{\'o}pez~Zunini, J.-P. Thivierge, S.~Kousaie, C.~Sheppard, and V.~Taler,
  ``Alterations in resting-state activity relate to performance in a verbal
  recognition task,'' \emph{PLoS One}, vol.~8, no.~6, p. e65608, 2013.

\bibitem{massimini2004sleep}
M.~Massimini, R.~Huber, F.~Ferrarelli, S.~Hill, and G.~Tononi, ``The sleep slow
  oscillation as a traveling wave,'' \emph{J. Neurosci.}, vol.~24, no.~31, pp.
  6862--6870, 2004.

\bibitem{scheeringa2012eeg}
R.~Scheeringa, K.~M. Petersson, A.~Kleinschmidt, O.~Jensen, and M.~C.
  Bastiaansen, ``{EEG} alpha power modulation of f{MRI} resting-state
  connectivity,'' \emph{Brain Connect.}, vol.~2, no.~5, pp. 254--264, 2012.

\bibitem{forbes2015spontaneous}
C.~E. Forbes \emph{et~al.}, ``Spontaneous default mode network phase-locking
  moderates performance perceptions under stereotype threat,'' \emph{Soc. Cogn.
  Affect. Neurosci.}, vol.~10, no.~7, pp. 994--1002, 2015.

\bibitem{sauseng2005fronto}
P.~Sauseng, W.~Klimesch, M.~Schabus, and M.~Doppelmayr, ``Fronto-parietal {EEG}
  coherence in theta and upper alpha reflect central executive functions of
  working memory,'' \emph{Int. J. Psychophysiol.}, vol.~57, no.~2, pp. 97--103,
  2005.

\bibitem{thung2018conversion}
K.-H. Thung \emph{et~al.}, ``Conversion and time-to-conversion predictions of
  mild cognitive impairment using low-rank affinity pursuit denoising and
  matrix completion,'' \emph{Med. Image Anal.}, vol.~45, pp. 68--82, 2018.

\bibitem{kim2019subject}
K.-T. Kim, C.~Guan, and S.-W. Lee, ``A subject-transfer framework based on
  single-trial {EMG} analysis using convolutional neural networks,'' \emph{IEEE
  Trans. Neural Syst. Rehabil. Eng.}, vol.~28, no.~1, pp. 94--103, 2019.

\end{thebibliography}

\end{document}